\documentclass[12pt]{article}
\pdfoutput=1
\usepackage{graphics}
\usepackage{graphicx, subfigure}
\usepackage{amssymb}
\usepackage{amsmath}
\usepackage[left=1.2in,right=1.2in,top=1.2in,bottom=1.2in]{geometry}
\usepackage{mathtools}

\newcommand{\rf}[1]{(\ref{#1})}
\newcommand{\beq}{\begin{equation}}
\newcommand{\eeq}{\end{equation}}
\newcommand{\bea}{\begin{eqnarray}}
\newcommand{\eea}{\end{eqnarray}}

\newcommand{\e}{\mbox{e}}

\newcommand{\g}{\gamma}

\renewcommand{\a}{\alpha}

\newcommand{\m}{\mu}

\newcommand{\del}{\delta}
\newcommand{\Del}{\Delta}

\newcommand{\oh}{\frac{1}{2}}

\newcommand{\ra}{\rangle}
\newcommand{\la}{\langle}

\newcommand{\cD}{{\cal D}}

\newcommand{\cM}{{\cal M}}

\begin{document}

\begin{center}
\vspace{24pt}
{ \large \bf Semi-classical Dynamical Triangulations}

\vspace{30pt}

{\sl J. Ambj\o rn}$\,^{a,b}$
and {\sl T.G. Budd}$\,^{a}$

\vspace{48pt}
{\footnotesize

$^a$~The Niels Bohr Institute, Copenhagen University\\
Blegdamsvej 17, DK-2100 Copenhagen \O , Denmark.\\
{ email: ambjorn@nbi.dk, budd@nbi.dk}\\

\vspace{10pt}
$^b$~Institute for Mathematics, Astrophysics and Particle Physics (IMAPP)\\ 
Radbaud University Nijmegen, Heyendaalseweg 135,
6525 AJ, Nijmegen, The Netherlands 
}
\vspace{36pt}
\end{center}


\begin{center}
{\bf Abstract}
\end{center}
\noindent
For non-critical string theory the partition function
reduces to an integral over moduli space after 
integrating over matter fields. The moduli integrand
is known analytically for genus one surfaces.
The formalism of dynamical triangulations  provides
us with a regularization of non-critical string theory
and we show that even for very small triangulations
it reproduces very well the continuum integrand when 
the central charge $c$ of the matter fields is large negative,
thus providing a striking example of how the quantum 
fluctuations of geometry disappear when $c \to -\infty$.

\vspace{12pt}
\noindent

\vspace{24pt}

\newpage

\section{Introduction}\label{intro}

Non-critical string theory (or equivalently 2d Euclidean quantum gravity
coupled to conformal field theories) has allowed us to study 
non-perturbative aspects of string theory and quantum gravity.
Certain aspects of the theory can be solved by continuum methods and 
certain aspects have been solved using combinatorial methods (matrix models
and dynamical triangulations (DT)). Originally  
dynamical triangulations were considered as a lattice regularization 
of path integral over two dimensional world sheet 
geometries \cite{david,kkm,adf}, 
but the versatility of the matrix models allowed for many other
interesting interpretations, which became increasingly 
detached from the original idea of DT as a lattice regularization
of a 2d quantum world. 

A typical dynamical triangulation represents a (piecewise linear) 
geometry which appears in the path  integral over 2d geometries. 
As such it does not represent anything like a physical measurable 
geometry, in the same way as a random path in the path integral 
representation of the particle is not physical and cannot be measured.
However, it has interesting fractal properties, precise as the random
path for the particle with probability 1 has Hausdorff dimension 2.
The ``random geometries'' in the path integral (represented in DT by 
dynamical triangulations) have Hausdorff dimension
\beq\label{1.0}
d_h= 2 \frac{\sqrt{49-c} +\sqrt{25-c}}{\sqrt{25-c} +\sqrt{1-c}}.  
\eeq
Contrary to the particle case, this is an {\it intrinsic} fractal
dimension, not  a fractal dimension in target space. Also,
it depends  on the central charge of conformal field theory 
coupled to 2d gravity. This result was first obtained for $c=0$ 
in \cite{kk} using the DT formalism, and later, again using 
the DT formalism, from a different perspective in \cite{fractal,fractal1}. 
The general formula was derived in \cite{watabiki} using 
Liouville theory.

\begin{figure}[t]
    \begin{center}
        \includegraphics[width=0.4\textwidth]{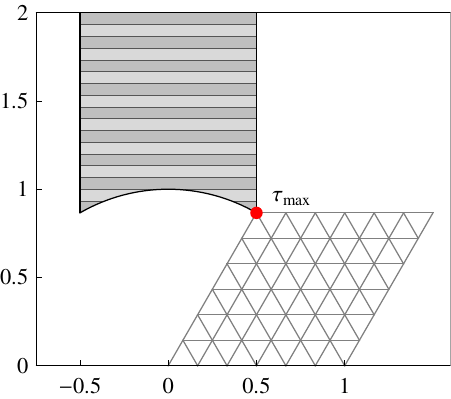}
    \end{center}
    \caption{The fundamental domain $\mathcal{M}$ in the upper-half complex plane. Its colloring illustrates the choice of bins used in Fig.\ \ref{fig1}. The maximum of $F(\tau)^{1-c}$ at $\tau_{\textrm{max}}=e^{i\pi/3}$ corresponds to the equilateral torus, i.e. the quotient of the plane by the hexagonal lattice. }
    \label{fig0}
\end{figure}

These results indicate that for a given central charge $c$ 
there exists a  measure defined on the set of continuous 
geometries. However, no precise mathematical definition exists yet,
contrary to the situation for continuous paths (the Wiener measure).
In critical  string theory this problem is circumvented
by using the conformal invariance of the theory. In this way 
most of the integration over geometries can be factored out,
and we are left only with a finite dimensional  integration over the moduli 
parameters (and a non-trivial integrand for higher genus surfaces). 
In non-critical string theory the situation is more complicated
due to the zero mode of the Liouville field, except in the genus 
one case. In this case one obtains just the 
standard result from critical string 
theory \cite{genus1}. Thus, if we consider the situation
where we fix the world sheet area to be $A$ we obtain
\beq\label{3s}
Z^{(h=1)}(A) \sim A^{-1}\int_\cM \frac{d^2\tau}{\tau_2^2}\; F(\tau)^{c-1}.
\eeq   
where $\tau=\tau_1+i \tau_2$ is the moduli parameter, $\cM$ denotes 
the fundamental domain of $\tau$ in the upper complex plane\footnote{
A convenient choice for the fundamental domain $\cM$ is (see Fig.\ \ref{fig0}):
\beq\label{31s}
\tau \in \cM ~~{\rm if}~~~
\left\{\begin{array}{lll} \tau_2 >0,~~-\oh < \tau_1 < 0 & ~~{\rm and} & 
~~|\tau|> 1 \\
 \tau_2 >0,~~~~~~ 0 \leq \tau_1 \leq \oh & ~~{\rm and} & 
~~|\tau|\geq 1 
\end{array}\right.
\eeq}
 and 
\beq\label{4s}
F(\tau) = \tau_2^{-1/2}|\eta(\tau)|^{-2}= 
\tau_2^{-1/2} \e^{\pi \tau_2/6} \prod_{n=1}^{\infty} 
|1-\e^{2\pi i n \tau}|^{-2}.
\eeq

The integral in \rf{3s} diverges for $c>1$, which tells us that there is a $c=1$ barrier beyond which the formulas make little sense. When $c$ decreases geometries with large $\tau_2$ become more and more suppressed.
In fact the maximum of $F(\tau)^{1-c}$, located at $\tau_{\mathrm{max}}=e^{i\pi/3}$ in the fundamental domain, will dominate the integral for $c\to -\infty$ (see Fig.\ \ref{fig0}). 

By assigning moduli parameters to triangulations of the torus one can 
study how the formalism of DT reproduces the {\it integrand} in \rf{3s}. 
The possibility to assign a moduli parameter to a toroidal 
triangulation was first realized in \cite{kawaigenus}, where the 
first study the DT-realization of eq.\ \rf{3s} was performed.  Using 
the formalism of simplicial cohomology this assignment can be done in a 
natural and relatively simple way \cite{abb}, and one can even 
explicitly construct the harmonic map from the geometry 
defined by the triangulation
to the parallelogram defined by the moduli parameter.
Stated briefly, given two independent cycles on a torus, $\g_1$ and 
$\g_2$, one can find dual harmonic one-forms $\a_1$ and $\a_2$ such that 
\beq\label{8s}
\int_{\g_i} \a^j = \del_i^j.
\eeq
Then a modular parameter of the torus is given by 
\beq\label{9s}
\tau = - \frac{\la \a^1 | \a^2\ra}{ \la \a^2 | \a^2\ra}
+i \sqrt{\frac{\la \a^1 | \a^1\ra}{ \la \a^2 | \a^2\ra} -
\left(\frac{\la \a^1 | \a^2\ra}{ \la \a^2 | \a^2\ra}\right)^2},
\eeq
where $\la \cdot, \cdot \ra$ refers to the standard inner product on one-forms.
This $\tau$ is not necessarily in the fundamental domain $\cM$,
but is uniquely associated with a point there by a modular transformation.
The concepts of cycles, one-forms and harmonic one-forms have 
a natural formulation on complexes,
the de Rham cohomology being replaced by simplicial cohomology.
In the case of DT, where the triangles are 
equilateral, all formulas become very simple.
On toroidal triangulations we can first find cycles
(non-contractible link-loops), then the dual one-forms (harmonic in 
the simplicial context), the corresponding $\tau$ and finally the 
modular transformation which maps it to the fundamental domain. 
We refer to \cite{abb} for details. 
 
In \cite{abb,abb1} this formalism was used to study 2d Euclidean 
quantum gravity for the matter central charges $c=0$ and $c=-2$. These 
values were chosen since they allowed for large computer generated 
triangulations with the correct weights. Large triangulations were necessary in order to properly include the singular geometries with large $\tau_2$.
In this article we extend the study to large negative $c$. For such $c$ we expect geometries with large $\tau_2$ to be suppressed and therefore smaller triangulations should suffice.
As we will explain below, relatively small triangulations {\it can}
be computer generated with the correct weight for any $c$.

\section{Large negative valued central charge}

Consider the partition function  for the bosonic string
in $d$ dimensions:
\beq\label{2.0}
Z^{(h)} = \int \cD [g] 
\int \cD_g X_\m \;\e^{-S(X,g)}, 
\eeq
where the integration is over worldsheet geometries $[g]$
of genus $h$ and Gaussian matter fields $X_\m$. The $d$ Gaussian 
fields can be integrated out and we obtain
\beq\label{1b}
\int \cD_{g} X_\m \; e^{-S(X,g)} \sim \Big(\det(-\Del_g')\Big)^{-d/2}.
\eeq 
Here $\Del'_g$ denotes the Laplace-Beltrami operator on the background
geometry defined by a the metric $g_{ab}$ 
and the prime signifies that the zero mode 
has been removed when calculating the determinant of $\Del_g$. If we 
consider $d$ as a formal parameter we can 
continue this expression to negative $d$. For positive integer $d$ we can 
identify $d$ with the central charge $c$ of a conformal field 
theory coupled to the 2d world sheet geometry and formally 
we can do the same for any real $d$. 
 The DT-formalism
tells us that we should represent the regularized partition function
\rf{2.0} as
\beq\label{2b}
Z^{(h=1)}(N) = \sum_{T_N}  \frac{1}{C_{T_N}}\; 
\Big({\det} (-\Del_{T_N}')\Big)^{-d/2},
\eeq
where the summation is over all triangulations of the torus with $N$
triangles. $C_T$ is the symmetry factor of the triangulation $T$, i.e.\ 
the order of the automorphism group of the graph $T$.
Finally $\Del_T$ denotes the (discretely defined) Laplacian on 
the DT-surface, which we take to be the usual graph Laplacian of 
the $\phi^3$ graph dual to the triangulation. 
The continuum area $A$ in formula \rf{3s} is related 
to $N$ by $A = N \,a^2 \sqrt{3}/2$, where $a$ denotes
the length of a link in the triangulation constructed from 
equilateral triangles. The continuum limit of DT is 
obtained by taking $a\to 0$ while keeping $A$ fixed. Thus we only
expect the DT result to be exact (i.e.\ to agree with the continuum 
expression) for $N\to \infty$.

\begin{figure}[t]
    \begin{center}
        \includegraphics[width=\textwidth]{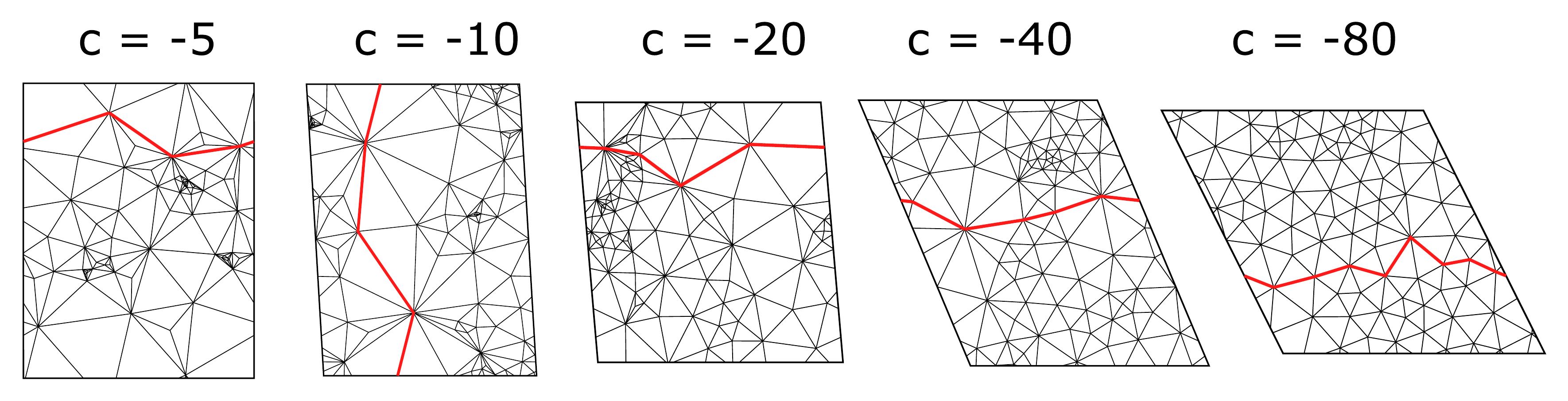}
    \end{center}
    
    \vspace{-0.5cm}
    \caption{Typical triangulations appearing in the partition function 
with $N=160$ triangles and increasingly negative central charge $c$. 
Thick (red) curves correspond to the shortest non-contractible 
loops studied in section \ref{sec3}.}
    \label{fig1a}
\end{figure}

As long as $N$ is not much larger than a few hundred triangles 
standard Monte Carlo simulations can be used to generate an ensemble of 
triangulations $\{ T_i(N)\}$ with the correct weight dictated by the 
partition function \rf{2b}. We refer to the Appendix for details 
on the updating algorithm used in the simulations. 
For each triangulation $T_i(N)$ we can now extract the moduli parameter
by the technique mentioned above and described in detail in \cite{abb}.
As mentioned before one can obtain an explicit harmonic map from the 
triangulation to the parallelogram in the Euclidean plane 
defined by the moduli parameters. 
Fig.\ \ref{fig1a} shows examples of such maps for typical 
triangulations appearing in the partition functions for various valus 
of the central charge. These examples indicate that typical 
triangulations become more and more regular, i.e. closer to the 
hexagonal lattice in Fig.\ \ref{fig0}, when $c$ becomes more negative.
From the measurements we construct a probability distribution in the 
fundamental domain $\mathcal{M}$ which we can compare with 
the continuum integrand $F(\tau)^{c-1}/\tau_2^2$ in \rf{3s}. 
The errors one encounters are twofold. There will be statistical errors 
which diminish with the size of the computer generated ensemble 
$\{ T_N\}$ in the standard way, and there will be systematic errors 
associated with the use a finite number $N$ of triangles.

\begin{figure}[t]
    \begin{center}
        \includegraphics[width=\textwidth]{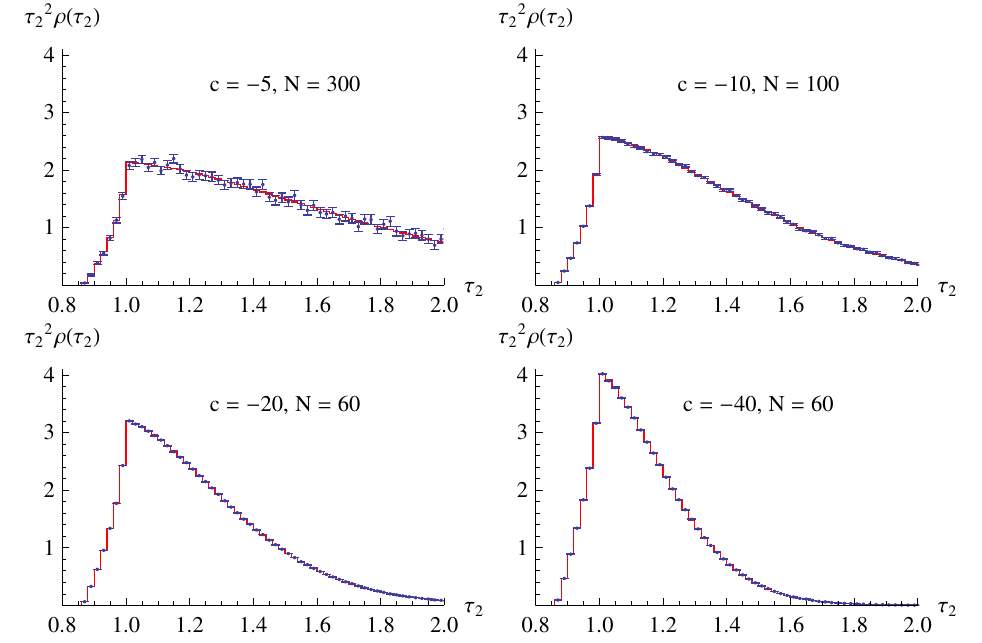}
    \end{center}
    \caption{The $\tau_2$ distributions $\tau_2^2\rho(\tau_2;N)$ 
    compared to the theoretical distributions. 
The data points and the theoretical curves are 
produced as described in the main text.}
    \label{fig1}
\end{figure}

Our results are shown in Fig.\ \ref{fig1}. The figures are constructed 
in the following way. The fundamental domain $\mathcal{M}$ is partitioned 
into bins according to the value of $\tau_2$ with bin size 
$\Delta\tau_2=0.025$, as illustrated in Fig.\ \ref{fig0} 
(however notice that there we have displayed bins with larger bin size 
for visualization purposes). We have chosen to disregard the value of 
$\tau_1$ since the theoretical integrand $F(\tau)^{c-1}/\tau_2^2$ 
varies little with $\tau_1$. We denote by $\rho(\tau_2)\Delta\tau_2$ 
the probability of observing a $\tau(T_i(N))$ associated with a 
triangulation $T_i(N)$ sitting in the bin 
$\tau_2-\oh\Delta\tau_2 \leq \tau_2(T_i(N)) < \tau_2+\oh\Delta\tau_2$. 
We compare $\rho(\tau_2)$ to the theoretical distribution 
$\rho_{\textrm{theory}}(\tau_2)$ obtained by integrating 
$F(\tau)^{c-1}/\tau_2^2$ over the corresponding bin. 
The measured distributions for various values of the central 
charge are depicted in Fig.\ \ref{fig1} by points with error-bars 
(which are hardly visible for $c \leq -10$) and the solid lines 
correspond to the theoretical distributions.



From Fig.\ \ref{fig1} it is seen that $\tau_2^2 \rho(\tau_2,N)$ is  
more and more concentrated at the smallest possible $\tau_2$ values when 
$c$ becomes increasingly negative, as expected from the theoretical 
distribution. Also, in order to obtain a reasonable accuracy 
we have to take $N=300$ for $c=-5$, while for $c=-40$ one obtains good 
results already with $N=60$. The need for large $N$ the closer $c$
is to barrier $c=1$ reflects that the distribution 
$\rho(\tau_2)$ becomes more flat when $c$ increases.
Thus both $\la \tau_2 \ra$ and  $\la \tau_2^2\ra -\la \tau_2\ra^2$ 
increases. However, the large $\tau_2$ region is 
not well reproduced by triangulations with small $N$,
the reason being that loops take 
integer length and, loosely speaking, $\tau_2$ is related to 
the ratio between the two independent shortest non-contractible loops.
When we get closer to $c=1$, the effective area of the fundamental
domain $\cM$ that has to be covered by the simulations becomes larger. 
Because of this and the need for larger triangulations, more and more 
computer-time is needed. This trend is reflected well
in the figures and the message is: the semi-classical region $c\to -\infty$
is easy!   

As further evidence of the precision with which the simplicial 
cohomology is able to reproduce the continuum results 
in the range of central charges that we have considered here,
we have fit the data in Fig.\ \ref{fig1} to the theoretical 
distribution with the central charge $c$ in formula \rf{3s} as a 
free parameter. The result is listed in Table \ref{table1}.


\begin{table}[ht]
        \begin{center}
                \begin{tabular}{|l||l|l|l|l|}

\hline
$c$   & $ -5$  & $-10$      & $-20$    & $-40$ \\ \hline
$c_{fit}$  & $ -5.01 \pm0.1$ & $-9.94 \pm 0.04$      & $-19.9 \pm 0.1$    
& $-39.88 \pm 0.05$ \\ \hline
                \end{tabular}
        \end{center}
        \caption{The fit $c_{fit}$ to the integrand in \rf{3s} from data.}
        \label{table1}
\end{table}

\section{The Hausdorff dimension}\label{sec3}

One of the most remarkable formulas in 2d Euclidean
quantum gravity is \rf{1.0}. Why do we need to verify it?
There have been a number of competing formulas \cite{competing}. 
Although it is mostly believed today that these formulas describe critical 
properties of the matter theories coupled to 2d gravity
rather than the fractal structure of a typical geometry
which appears in the path integral \cite{akw}, the question 
is not completely settled. In addition,
the derivation of \rf{1.0} is based on a certain number
of assumptions related to diffusion on fluctuating geometries,
assumptions which are not necessarily valid.   

In \cite{abb,abb1} formula \rf{1.0} 
was verified with great numerical accuracy for $c=0$ and $c=-2$.
Here we check the formula for more negative $c$. As noted in 
\cite{abb,abb1} the torus offers a nice opportunity 
for a numerical check of the formula which is independent of the original 
numerical check \cite{hausdorff} 
performed by simply measuring the area $A(r)$ enclosed 
in disks  of geodesic radius $r$:
\beq\label{3.0}
\la A(r) \ra \sim r^{d_h},~~~~~r \ll A^{1/d_h},
\eeq
where $A$ is the fixed area of the 2d geometry, as in formula \rf{3s}.
On the torus one can use that in a given geometry
a shortest non-contractible loop must be a geodesic curve. 
On average a non-contractible loop will have a length 
$\la L \ra_A \sim A^{1/2}$. However, if the geodesic distances scale
anomalously, as is the case if $d_h \neq 2$ in \rf{3.0},
we expect instead
\beq\label{3.1}
\la L \ra_A \sim A^{1/d_h}.
\eeq

\begin{figure}[t]
\vspace{-0.5cm}
\begin{center}
\includegraphics[width=\textwidth]{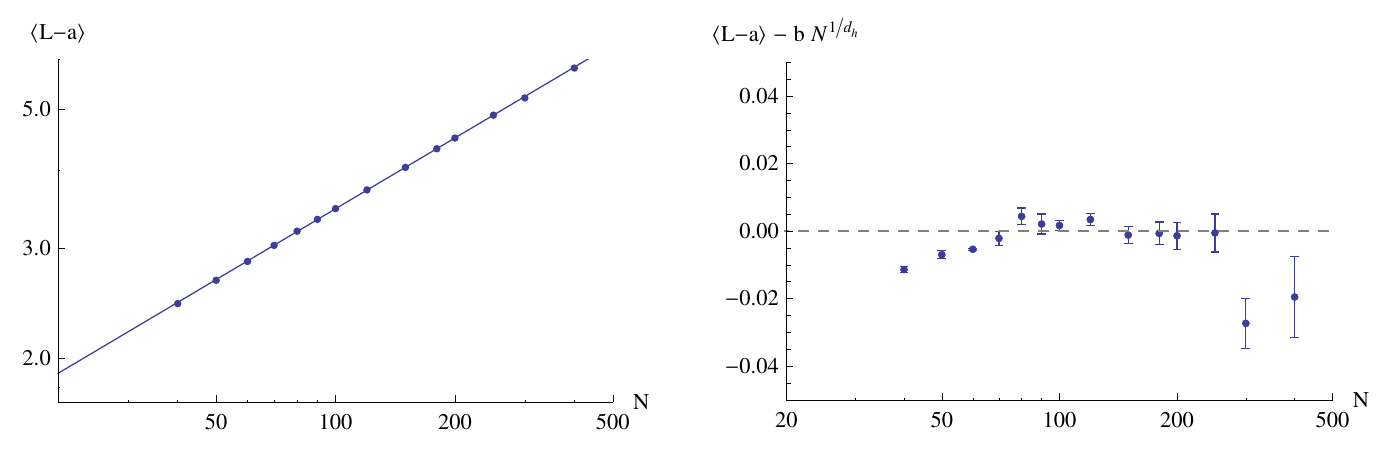}
\end{center}
\caption{Left figure: a log-log plot of $\la (L-a) \ra_N$ 
versus $N$ as well as 
the theoretical curve $ b \,N^{1/d_h}$ for 
$c=-20$, $a=-0.10, b=0.61$ determined 
by a best fit. The range of $N$ is 30--400. 
Right figure: the difference $\la (L-a) \ra_N
- b \,N^{1/d_h}$.}
\label{fig2}
\end{figure}

In DT we can perform such measurements: we define curves as link-paths,
and a geodesic curve between two vertices as a shortest link-path. Let 
now $\{T_i(N)\}$ denote the DT ensemble of triangulations with 
$N$ triangles generated 
with the correct weight. Fixed $N$ corresponds to fixed continuum area 
$A$ in this context. For each triangulation we determine a shortest 
non-contractible loop\footnote{The practical algorithm used to 
find the shortest non-contractible loop uses in a nice way
simplicial cohomology. We refer to \cite{abb} for details. 
See also Fig.\ \ref{fig1a} for examples of shortest non-contractible loops.}
and in this way we can calculated the average $\la L \ra_N$. 
We do this for as large range of $N$ as possible and determine $d_h$
from a fit to \rf{3.1}. We have chosen $c=-20$ and $N$ in the range from
30 to 400. A best three parameter fit\footnote{The parameter $a$ in 
\rf{3.2} is not present in \rf{3.1}, but is quite natural. It 
reflects the fact that $L$ in DT is an integer, contrary to 
the $L$ in the continuum formula. It can be viewed as 
a finite size correction, using the terminology of finite size 
scaling in the theory of critical phenomena.} 
\beq\label{3.2} 
\la (L-a) \ra_N = b \; N^{1/d}  ~~~\rightarrow ~~~ d = 2.76\pm 0.07,
\eeq
which should be compared to the theoretical value $d_h \approx 2.66$.
In view of the small values of $N$ used
we consider this result as quite satisfactory and a confirmation
of formula \rf{1.0}. The left figure in Fig.\ \ref{fig2} shows 
the measured $\la (L-a) \ra_N$ together with the theoretical 
curve $b \, N^{1/d_h}$ (where $a,b$ is determined by a best fit).
The right figure in Fig.\ \ref{fig2} displays the error in the fit.   
It is clear that $N$ is too small to produce a perfect fit. If one 
leaves out the three smallest values of $N$ one obtains a $\chi^2$ 
value of order 2, which is good.

\section{Discussion}
 
Matrix models have provided convincing evidence that the formalism
of dynamical triangulations provides a regularization of non-critical 
string theory or equivalently 2d Euclidean quantum gravity coupled 
to matter with central charge $c\leq 1$. However, quantum {\it geometric}
aspects of 2d Euclidean quantum gravity are not easily analyzed using 
matrix models. The reason is that when using matrix models  one 
integrates over all geometries. The matrix model thus provides 
us with the full partition function of matter and gravity and to 
obtain more detailed information one has to insert certain ``punctures''
and calculate their expectation values. These operators also mix 
matter and geometry and to disentangle geometry and matter is not 
easy and becomes increasing complicated in the limit $c\to -\infty$,
a limit where one naively would expect the quantum aspects of 
geometry to decouple. Formula \rf{1.0} can be read as 
an explicit, quantitative statement about this decoupling: $d_h \to 2$
for $c \to -\infty$. However, it is not derived using matrix models but 
quantum Liouville theory.

One would expect that the original interpretation of DT, where the 
triangulations are viewed as a lattice regularization of 
the path integral over geometries, should allow us to observe
the increased semi-classical nature of geometry in detail. The 
purpose of this article was to provide evidence for this. 
For the torus we have shown that the formalism of DT indeed 
reproduces the {\it integrand} in \rf{3s}, not only the 
partition function. Further, the more negative the value
of $c$, the better is the continuum expression reproduced
even for small triangulations. While conformal invariance of 
the continuum theory  blurs somewhat the concept of a 
{\it regular} geometry, the possibility to use the moduli
parameter of the torus has allowed us to address the problem
in a conformally invariant way, and in the $c\to -\infty$ limit 
we clearly see the dominance of the ``regular'' triangulations, 
i.e. triangulations of the torus which have small $\tau_2$ values.

Finally, using the concept of a shortest  non-contractible 
loop on the torus, we have been able to test and (to some degree)
verify formula \rf{1.0}.

\subsection*{Acknowledgment}\label{app}

The authors acknowledge support from the ERC-Advance grant 291092,
``Exploring the Quantum Universe'' (EQU). JA acknowledges support 
of FNU, the Free Danish Research Council, from the grant 
``quantum gravity and the role of black holes''.
Part of this research was carried out at Utrecht University, The Netherlands.

\section*{Appendix: Monte Carlo simulations}

\begin{figure}[t]
    \begin{center}
        \includegraphics[width=0.4\textwidth]{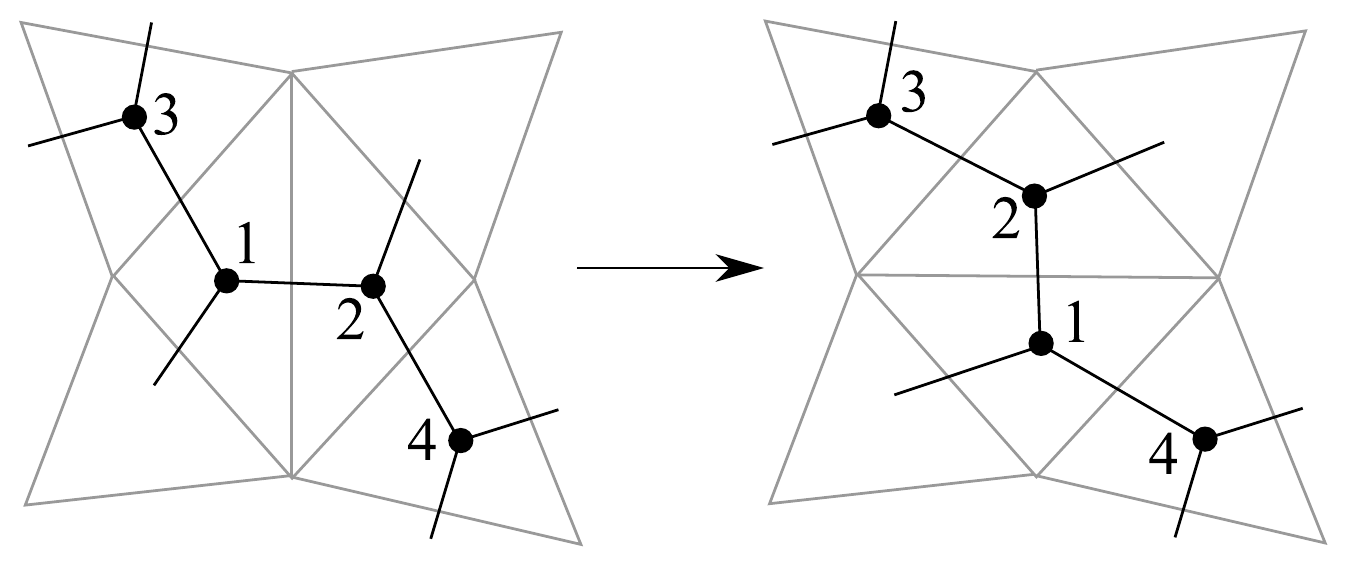}
    \end{center}
    \caption{Flip move.}
    \label{figflip}
\end{figure}

In order to generate random triangulations according to the distribution \rf{2b} we use standard Monte Carlo methods. We start with some initial triangulation with the desired number $N$ of triangles. Then we apply random updates to the triangulation via so-called flip moves, as shown in Fig.\ \ref{figflip}. To obtain the correct Boltzmann distribution one has to implement an acceptance probability that is related to the ratio of Boltzmann factors before and after the flip move, i.e. 
\beq\label{A.1}
p=\left[\frac{\det\nolimits'(-\Delta-\delta\Delta)}{\det\nolimits'(-\Delta)}\right]^{-d/2} = \left[\det(I + \Delta^{-1}\delta\Delta)\right]^{-d/2}.
\eeq
Here $\delta\Delta$ is the change under the flip move of the Laplacian on the dual graph of the triangulation, and $\Delta^{-1}$ is the pseudo-inverse of $\Delta$. The matrix $\delta\Delta$ has only eight non-vanishing entries and, adopting the labeling in Fig.\ \ref{figflip}, is given by
\beq
\delta\Delta = \begin{pmatrix*}[r] 
0 & 0 & -1 & 1 & \cdots \\
0 & 0 & 1 & -1 & \cdots \\
-1 & 1 & 0 & 0 & \cdots \\
1 & -1 & 0 & 0 & \cdots \\
\vdots & \vdots & \vdots & \vdots & \ddots
\end{pmatrix*}.
\eeq
Therefore \rf{A.1} reduces to a determinant of a four by four matrix
\beq
p=\begin{vmatrix*}[r]
1-\psi_1 & \psi_1 & -\phi_1 & \phi_1 \\
\psi_1 & 1-\psi_1 & \phi_1 & -\phi_1 \\
-\psi_3 & \psi_3 & 1-\phi_3 & \phi_3 \\
\psi_3 & -\psi_3 & \phi_3 & 1-\phi_3 
\end{vmatrix*}^{-d/2},
\eeq
where the vectors $\phi$ and $\psi$ are the solutions to the linear equations
\beq
\Delta \phi = \begin{pmatrix*}[r] 1 \\ -1 \\ 0 \\ 0 \\ \vdots\end{pmatrix*}, \quad \Delta \psi = \begin{pmatrix*}[r] 0 \\ 0 \\ 1 \\ -1 \\ \vdots\end{pmatrix*}, \quad \sum_i \phi_i =\sum_i \psi_i = 0.
\eeq
Solving these equations using numerical linear algebra techniques is the most time-consuming step of the update move. It limits the simulations to triangulations of up to several hundred triangles.

\end{document}